\documentclass[a4paper]{article}

\usepackage{INTERSPEECH2015}

\usepackage{graphicx}
\usepackage{amssymb,amsmath,bm}
\usepackage{textcomp}
\usepackage{multirow}
\usepackage{url}
\usepackage{slashbox}

\ninept

\title{Contaminated speech training methods for robust DNN-HMM \\distant speech recognition}

\makeatletter
\def\name#1{\gdef\@name{#1\\}}
\makeatother \name{{\em Mirco Ravanelli, Maurizio Omologo}}

\address{Fondazione Bruno Kessler, Trento, Italy \\
  {\small \tt mravanelli@fbk.eu, omologo@fbk.eu}
}

\begin{document}

  \maketitle
  \begin{abstract}
Despite the significant progress made in the last years, state-of-the-art speech recognition technologies provide a satisfactory performance only in the close-talking condition. Robustness of distant speech recognition in adverse acoustic conditions, on the other hand, remains a crucial open issue for future applications of human-machine interaction. To this end, several advances in speech enhancement, acoustic scene analysis as well as acoustic modeling, have recently contributed to improve the state-of-the-art in the field.
One of the most effective approaches to derive a robust acoustic modeling is based on using contaminated speech, which proved helpful in reducing the acoustic mismatch between training and testing conditions.

In this paper, we revise this classical approach in the context of modern DNN-HMM systems, and propose the adoption of three methods, namely, asymmetric context windowing, close-talk based supervision, and close-talk based pre-training. The experimental results, obtained using both real and simulated data, show a significant advantage in using these three methods, overall providing a 15\% error rate reduction compared to the baseline systems. The same trend in performance is confirmed either using a high-quality training set of small size, and a large one.

  \end{abstract}

  \section{Introduction} \label{sec:intro}
  During the last decade, much research has been devoted to improving Automatic Speech Recognition (ASR) performance. 
Nevertheless, most state-of-the-art systems are still based on close-talking solutions, forcing the user to speak very close to a microphone-equipped device. 
There are, however, various real-life situations where  Distant Speech Recognition (DSR) is more natural, convenient and attractive \cite{dasr}.
An emerging application is, for instance, speech-based domestic control, where users might prefer to freely interact with their home appliances without wearing or even handling any microphone-equipped device. This scenario was addressed under the EU DIRHA project\footnote{This work was partially funded by EU's 7th Framework Programme under grant agreement n. 288121 DIRHA. More details can be found in \url{http://dirha.fbk.eu/}.}, which had the ultimate goal of developing real-time systems for motor-impaired users. 
Unfortunately, 
current DSR technologies still exhibit a significant lack of robustness and flexibility, due to the adverse acoustic conditions originated by non-stationary noises and acoustic reverberation \cite{adverse}. 

Several efforts have been devoted during the last years to improve DSR, as also witnessed by some recent international challenges such as REVERB \cite{revch}, CHIME \cite{chime,chime3} and ASpIRE. Considerable progresses were fostered by substantial advances in spatial filtering \cite{BrandWard,beam}, microphone selection \cite{nadeu}, source separation \cite{bss}, speech dereverberation \cite{derev} as well as speaker localization \cite{hscma}, acoustic event detection \cite{aed1} and acoustic modeling. 
Concerning the latter field, the research community is currently experiencing a revolution due to the introduction of Deep Neural Networks (DNNs) \cite{lideng}, which have consistently outperformed previous Gaussian Mixture Models (GMMs) for both close \cite{IEEEexample:intro1} and distant-talking scenarios \cite{pawel2,hain,dnn_rev,dnn_rev2,dnn3}. 
One of the most effective and straightforward approach so far applied to DSR for acoustic modeling is based on multi-condition training with contaminated speech corpora \cite{matassoni,cont2,cont3}. 
Contaminated speech is generated by convolving close-talking signals with an Impulse Response (IR) usually measured in the targeted environment. 
Background noise is then typically added to the convoluted signals, in order to make the resulting corpus more realistic and better matching real-world conditions. This approach has also been adopted recently in \cite{revch,chime,rav_in14,brutti}.


This paper deals with the use of contaminated speech to train a DNN-HMM speech recognition system. Training such very complex systems is typically difficult and tricky. The basic strategy to run it, and the way one performs optimization steps, often play a crucial role in order to get best performance. Our work, which relies on the Kaldi toolkit \cite{kaldi}, 
has concerned all those optimization efforts, e.g., including an optimal choice of number of hidden layers and of units. However, in this paper we focus on two specific aspects, about which we believe the experimental results demonstrate novel and interesting evidences.

First, we propose some methods able to exploit information from both distant and close-talking datasets to derive robust acoustic models. In particular, one proposed approach consists in inheriting the high-quality labels that might be generated from close-talk datasets to train distant-talking DNNs. Another novel method consists in replacing a standard Restricted Boltzmann Machines (RBMs) based pre-training \cite{rbm1} with a supervised close-talk pre-training, leading to a smarter initialization of the distant-talking DNN.

As a second contribution, we propose to replace the traditional symmetric context windows, which are typically adopted to gather several consecutive features in the DNN framework, with an Asymmetric Context Window (ACW). Interestingly, this approach has proved effective in counteracting the harmful effects of the acoustic reverberation, due to a better use of the contextual information. 


The experimental validation is carried out in a living-room of a real apartment using both simulated and real test data. 
Similarly to \cite{rav_in14}, the evaluation is performed on a phone-loop task, in order to avoid any non-linear influence and artifacts possibly originated by a language model.

The rest of the paper is organized as follows. In Sec. \ref{sec:cont} the contamination approach is outlined, while Sec. \ref{sec:corpora} and \ref{sec:ASR} describe the adopted corpora and the ASR system, respectively. Finally, the proposed methods are experimentally validated in Sec. \ref{sec:exp}, while Sec. \ref{sec:con} draws the conclusions.  


\section{Speech contamination} \label{sec:cont}

The speech contamination process is represented by the  following basic relationship:
 \begin{equation}
 y(t)=x(t)*h(t)+\alpha n(t)
 \label{eq:cont}
 \end{equation}
where $x(t)$ and $y(t)$ are the close and distant-talking signals, respectively, while $n(t)$ consists in recorded environmental noise controlled by the gain $\alpha$ in order to obtain different SNRs. $h(t)$ is an IR that describes the acoustic reverberation effects of a generic source-microphone path.
In particular, if one assumes to deal with a linear time-invariant acoustical transmission system, the IR provides a complete description of the changes a sound signal undergoes when it travels from a particular position in space to a given microphone \cite{kutt}. 
The method adopted here for estimating IR consists in diffusing in the target environment 
a known Exponential Sine Sweep (ESS) signal, and recording it by a microphone properly placed in space \cite{farina}.
This method of measuring the IR has a remarkable impact on the speech recognition performance, as discussed in \cite{ravanelli}.


\section{Simulated and Real Corpora} \label{sec:corpora}
The contamination process described in Sec. \ref{sec:cont}, can be exploited to generate simulated data starting from different close-talking datasets. 
In this work, simulated data are used for training and development purposes only, while test is performed also on real data. The sampling rate is 16 kHz for all the datasets. In the following sections, the reference environment, the close-talking corpora, and the simulated and real datasets are described. 
  
\subsection{Acoustic Environment and Microphone Set-up} \label{sec:env}
The reference environment considered for this study is the living-room of a real apartment, which was available for experiments under the DIRHA project.
\begin{figure}
\centering
\includegraphics[width=0.45\textwidth]{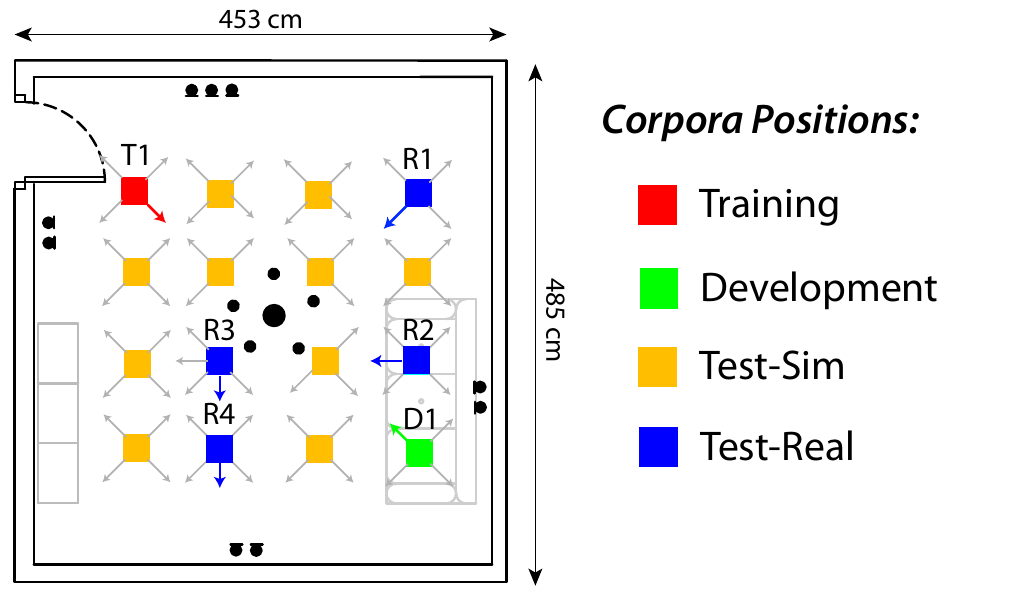}
\caption{The reference DIRHA living-room environment. Squares and arrows represent the speaker positions and orientations adopted for the real and simulated corpora. }
\label{fig:dirhaflat}
\end{figure}
The original microphone network includes several microphones installed on the walls and on the ceiling.
However, due to the purpose of this work, in the following we will consider only one microphone, i.e., the central one of the ceiling array.
As shown in Fig. \ref{fig:dirhaflat}, an IR measurement session exploring 
a large number of positions and orientations of the sound source has been conducted. 
The mean reverberation time $T_{60 }$ measured in the room is $0.7$ s. 

\subsection{Close-talking Corpora} \label{sec:ct}
In this work, two close-talk datasets are used for training, namely APASCI and Euronews. APASCI \cite{apasci1}  consists of more than 5200 phonetically-rich sentences uttered by 163 speakers and recorded in a professional studio, overall corresponding to about 6 hours. 
For comparison purposes, the Italian part of the Euronews database \cite{gretter}, consisting of  about 100 hours of TV news, is also considered for training. The Euronews signals are frequently corrupted by cross-talk, background music, environmental noise as well as spontaneous speech phenomena.

A small development set is also adopted to tune the free-parameters of the DSR systems, as discussed in Sec. \ref{sec:baselines}.
To this purpose, a portion of the DIRHA corpus \cite{lrec}, consisting of 220 high-quality phonetically-rich sentences uttered by 10 speakers is employed. 

For testing purposes, a different portion of the DIRHA corpus, composed of 430 phonetically-rich sentences uttered by 20 speakers is adopted. 

    

\subsection{Simulated Corpora} \label{sec:sim}
For each of the close-talking corpora outlined in Sec. \ref{sec:ct}, two contaminated versions were generated. The first version is based on a convolution of the close-talking sentences with impulse responses measured in the reference living-room environment. Such simulated datasets are intended to study a scenario where only reverberation acts as a source of disturbance. Furthermore, a second version  of the contaminated databases was directly derived from the reverberated datasets  by adding noisy background sequences recorded in the target environment, ensuring a mean SNR of about 10 dB . This version is intended to study more challenging scenarios where both reverberation and noises act as a source of disturbance.
For the training corpora a single impulse response (from T1 in Fig.\ref{fig:dirhaflat}) was considered, while for development a different IR (D1) was adopted.
As shown in \cite{rav_in14}, note that a single impulse response is sufficient to model room reverberation effects for DSR purposes. 

\subsection{DIRHA Real Corpus} \label{sec:real}
Besides the simulated test-sets, a corpus of real phonetically-rich sentences was recorded in the reference living-room. 
These real recordings involved 18 native Italian speakers (9 males and 9 females), who read a list of 40 phonetically-rich sentences in four different positions (R1-4 in Fig.\ref{fig:dirhaflat}). In total, 2880 sentences, corresponding to more than two hours of speech material were acquired. These recordings were conducted in quiet acoustic conditions, where reverberation was the main source of disturbance. The estimated SNR is about 22 $dB$. 


\begin{figure*}[t!]
  \includegraphics[width=\textwidth,height=4cm]{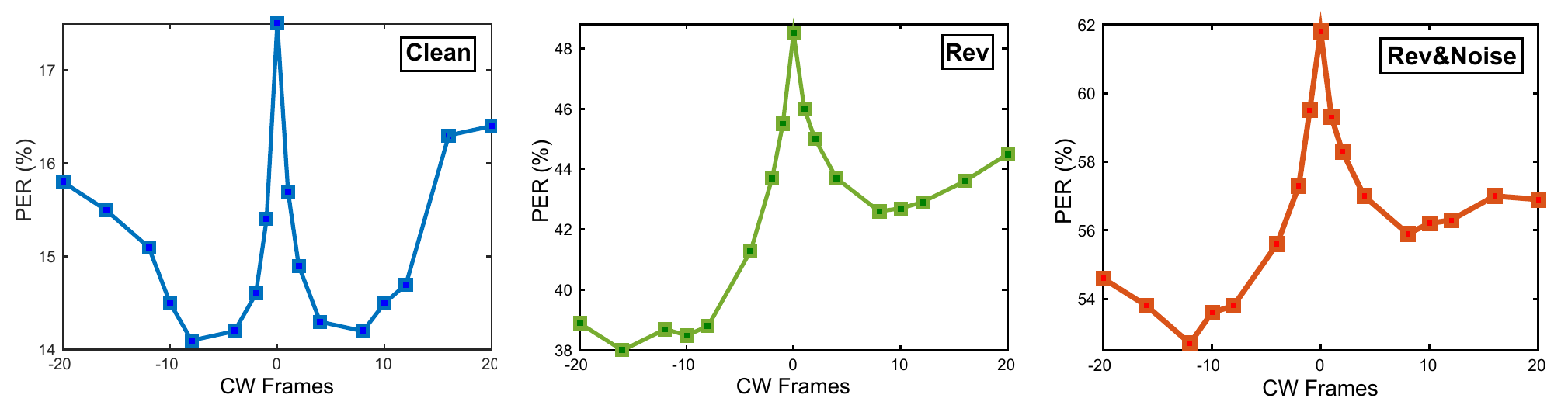}
  \caption{PER(\%) obtained by DNN context windows which progressively integrate past or future frames.} 
\label{fig:exp1}
\end{figure*}

\section{DNN based ASR} \label{sec:ASR}
In this work, we use a Context-Dependent DNN-HMM speech recognizer, where every unit is modeled by a three state left-to-right HMM, and the tied-state observation probabilities are estimated through a DNN. 

Feature extraction is based on blocking the signal into frames of 25 ms with 10 ms overlapping. For each frame, 13 MFCCs plus pitch and Probability of Voicing (PoV) are extracted. The pitch and PoV are estimated through the normalized autocorrelation method discussed in \cite{kpitch}. The resulting features, together with their first and second order derivatives, are then arranged into a single observation vector of 45 components. 
Finally, a context window gathering several consecutive frames followed by a mean and variance normalization of the feature space are applied before feeding the DNN. 

The DNN, trained with the Kaldi toolkit (Karel's recipe) \cite{kaldi}, is composed of sigmoid-based hidden neurons, while the output layer is based on softmax activation functions.
The pre-training phase is carried out by stacking Restricted Boltzmann Machines (RBM) \cite{rbm1} to form a deep belief network, while 
the fine-tuning is performed by a stochastic gradient descent optimizing cross-entropy loss function.
In the latter phase, the initial learning rate is kept fixed as long as the increment of the frame accuracy on the dev-set is higher than 0.5$\%$. For the following epochs, the learning rate is halved until the increment of frame accuracy is less than the
stopping threshold of 0.1$\%$. 

The decoding is performed 
by adopting a phone-loop based grammar.
As previously outlined, even though the use of more complex grammars or language models is certainly helpful in increasing the recognition performance, the adoption of a simple phone-loop is due to the need of an experimental evidence not biased by a LM.
In this study, as in \cite{rav_in14}, a set of 26 phone units of the Italian language was chosen for evaluation purposes.


  \section{Experiments} \label{sec:exp}
In the following section, the proposed techniques are better described and an experimental validation is provided\footnote{Part of the experiments are conducted with a Tesla K40 donated by the NVIDIA Corporation.}. The experiments are conducted in three different acoustic scenarios of increasing complexity: close-talking, distant-talking with reverberation (Rev), and distant-talking with both noise and reverberation (Rev\&Noise).
The reported performance is obtained when a matching acoustic condition between test and training is met (e.g., for the reverberated speech test, the reverberated training is used). The evaluation is performed on both simulated and real data, using either APASCI or Euronews training-sets.

  \subsection{Baselines} \label{sec:baselines}
This section reports the results obtained with a standard DNN-HMM system, after optimizing the main ASR free-parameters on the dev-set. The best performance is reached with a DNN composed of 6 hidden layers with 1500 neurons per layer, adopting a symmetric context window of 17 frames (8 before and 8 after the current frame) and using a learning rate of 0.008. 
 Note that such free-parameters are used for both the training corpora and for all the addressed acoustic scenarios, due to a similar outcome of the optimization step. Only a difference concerning the optimal number of tied-states is observed, leading to 2.5k and 5k tied-states for APASCI and Euronews, respectively.
Table \ref{tab:baselines} shows the performance achieved on the test-sets.
  

  \begin{table}[h!]
\centering
\small
    \begin{tabular}{ | l | c | c | c |}
    \hline
    \backslashbox{\em{Test}}{\em{Train}} & APASCI (6 h) & Euronews (100 h)   \\ \hline
    Close-talk & 14.0 & 14.5  \\ \hline
    Sim-Rev & 37.6 & 37.0  \\ \hline
    Real-Rev & 41.0 & 40.1  \\ \hline
    Sim-Rev\&Noise & 52.3 & 51.8  \\ \hline

    \end{tabular}
\caption{Phone Error Rate (PER\%) of the baseline system.}
\label{tab:baselines}
\end{table}

Table \ref{tab:baselines} clearly highlights that the performance in distant-talking scenarios is dramatically reduced, if compared to a close-talking case. This loss of performance further confirms how challenging DSR is under adverse acoustic conditions and without any help from grammars or LMs. It is also highly relevant that with only 6 hours of high-quality speech material a performance similar to that of a 100 hours corpus is obtained.
  
\subsection{Asymmetric Context Window (ACW)} \label{sec:acw}
In distant-talking scenarios, the acoustic reverberation introduces a long-term redundancy in the signal. This disturbance can be modeled as a long causal FIR filter, whose taps introduce a forward memory in the speech sequence. 
As a consequence, since the length of the window adopted for features extraction (i.e., 25 ms) is much smaller than the reverberation time (i.e, 700 ms), a progressive concatenation of future frames tends to embed information correlated with the central frame contents. On the other hand, a concatenation of past frames seems to be more helpful since, on average, different and complementary information can potentially be analyzed to perform a frame-level prediction. Based on this concept, we explore the use of an Asymmetric Context Window (ACW), which analyzes more past frames than future frames. 

Fig. \ref{fig:exp1} reports a first experiment involving context windows which progressively integrate only past (negative x-axis) or future frames (positive x-axis). In this section, only results obtained using APASCI as training-set are reported. However, a similar trend is observed with Euronews.

From Fig. \ref{fig:exp1}, it is clear that in a close-talking scenario a symmetric behaviour is attained, meaning that past and future information provides a similar contribution in improving the system performance. Conversely, the role of the past information seems significantly more important in distant-talking scenarios, since a faster decrease of the PER(\%) is progressively obtained when past frames are concatenated. 
Along this line, another experiment, reported in Table \ref{tab:test2}, shows the results achieved when the symmetric context window of 17 frames introduced in Sec. \ref{sec:baselines} is replaced by different asymmetric context windows.

\begin{table}[h!]
\centering
\small
\tabcolsep=0.11cm
    \begin{tabular}{ | l | c | c | c | c | c |}
    \hline
    \multirow{2}{*}{\backslashbox{\em{Test}}{\em{Train}}} & \multicolumn{5}{ | c | }{Context Window} \\  \cline{2-6}
    & P16-F0 & P10-F6 & P8-F8 & P6-F10 & P0-F16 \\ \hline
    Close-talk & 15.5 & 14.2 & \bf14.0 & 14.4 & 16.3 \\ \hline
    Sim-Rev & 38.0 & \bf37.0 & 37.6 & 38.3 & 43.6 \\ \hline
    Real-Rev & 41.6 & \bf40.2 & 41.0 & 41.6 & 46.8 \\ \hline
    Sim-Rev\&Noise & 53.8 & \bf51.8 & 52.3 & 53.0 & 57.0 \\ \hline
    
    \end{tabular}
\caption{PER(\%) obtained with Symmetric and Asymmetric context windows composed of 17 frames. ``P'' refers to the number of past frames, while ``F'' refers to future frames.}
\label{tab:test2}
\end{table}

Consistently with the previous experiment, in a close-talking scenario the typical symmetric window performs better than any asymmetric window. 
Differently, an asymmetric context window which embeds more past information (10 frames) than future information (6 frames) performs slightly better in distant-talking conditions. Note that this is consistent for both real and simulated test-sets. Moreover, results obtained with different context window lengths (e.g., 9, 11, 13, 21, 27 frames) further confirms that ACW is a better choice in distant-talking scenarios, thanks to an improved robustness against reverberation.  
Note also that the use of an ACW does not imply any additional computational effort and is particularly suitable for on-line systems \cite{online2,acw1}, since the latency originated by waiting for the future frames is minimized.


  \subsection{Close-talking labels for Distant-talking DNN} \label{sec:ct1}
In standard ASR, the labels for DNN training are derived by a forced-alignment of the training corpus over the tied-states. Although some GMM-free solutions have been proposed \cite{gmm-free}, this alignment is typically performed using a standard CD-GMM-HMM system. 
However, this phase can be very critical because a precise alignment could be difficult to reach, especially in challenging acoustic scenarios characterized by noise and reverberation. As a consequence, the DNN learning process might be more critical due to a poor supervision.   

In the contaminated speech training framework, however, a more precise supervision can be obtained from the close-talking dataset.
For this reason, we propose to inherit the high-quality labels that can be generated from the close-talking data to train the distant-talking DNNs. This requires to train, with the original clean datasets, a standard CD-GMM-HMM system, and exploit it to generate a precise tied-state forced alignment over the close-talking training corpus, later inherited as supervision for the distant-talking DNN.
Table \ref{tab:test3} reports the results obtained with the standard approach, which is based on labels derived from distant-talking signals, and the proposed solution, which directly inherits labels derived from close-talking signals. The ACW of 17 frames (P10-F6) discussed in Sec. \ref{sec:acw} is still used for the following experiments.



     

\begin{table}[h!]
\centering
\small
\tabcolsep=0.11cm
    \begin{tabular}{ | l | c | c | c | c | }
    \cline{1-5}
    \multirow{2}{*}{\backslashbox{\em{Test}}{\em{Train}}} & \multicolumn{2}{ | c |}{APASCI (6 h)}  & \multicolumn{2}{ | c |}{Euronews (100 h)}  \\ \cline{2-5}
    & Standard & CT-lab & Standard & CT-lab \\ \hline
    Sim-Rev & 37.0 & \bf33.0 & 36.1 & \bf32.1 \\ \hline 
    Real-Rev & 40.2 & \bf35.9 & 39.3 & \bf34.3 \\ \hline 
    Sim-Rev\&Noise & 51.8 & \bf47.3 & 50.1 & \bf46.4 \\ \hline      
    \end{tabular}
\caption{PER(\%) obtained in distant-talking scenarios with the standard and with the proposed technique based on close-talking labels (CT-lab).}
\label{tab:test3}
\end{table}

Results show that the proposed approach provides a substantial improvement in the performance, over both real and simulated data. Specifically, a relative improvement of 10\% and 12\% is achieved for APASCI and Euronews, respectively.
Besides this significant performance improvement, another interesting aspect is the faster convergence of the iterative learning procedure, due to a better supervision provided to the DNN. In these experiments, 15 epochs are needed to converge with the standard solution, while only 12 are sufficient with the proposed approach, reducing the training time of 20\%.

  \subsection{Close-talking Pre-Training for Distant-talking DNN} \label{sec:ct2}
In this section, we propose a further way of taking advantage of the rich information embedded in the close-talking dataset. In particular, instead of adopting a standard unsupervised RBM pre-training of the DNN, we propose to use a supervised pre-training method based on the close-talking data. More precisely, the idea is to train a close-talking DNN and inherit its parameters for initializing the distant-talking DNN. A subsequent fine tuning phase is then carried out on distant-talking data using a slightly reduced learning rate (LR=0.005 is used here). 
Table \ref{tab:test4} reports the results obtained using such approach on real and simulated datasets. The ACW (P10-F6) discussed in Sec. \ref{sec:acw} and the close-talking labels proposed in Sec. \ref{sec:ct1} are still used for this experiment.

\begin{table}[h!]
\centering
\tabcolsep=0.15cm
    \begin{tabular}{ | l | c | c | c | c | }
    \cline{1-5}
    \multirow{2}{*}{\backslashbox{\em{Test}}{\em{Train}}} & \multicolumn{2}{ | c |}{APASCI (6 h)}  & \multicolumn{2}{ | c |}{Euronews (100 h)}  \\ \cline{2-5}
    & Stand-PT & CT-PT & Stand-PT & CT-PT \\ \hline
    Sim-Rev & 33.0 & \bf31.4 & 32.1 &  \bf31.2 \\ \hline 
    Real-Rev & 35.9 &  \bf34.7 & 34.3 &  \bf33.0 \\ \hline 
    Sim-Rev\&Noise &  47.3 & \bf46.2 & 46.4 &  \bf45.1 \\ \hline

    \end{tabular}
\caption{PER(\%) obtained in distant-talking scenarios with a standard RBM pre-training (Stand-PT) and with the proposed supervised close-talking pre-training (CT-PT).}
\label{tab:test4}
\end{table}

Results clearly show an improvement deriving from this approach, which is consistent over both simulated and real data. 
This suggests that a close-talk pre-training is a smart way of initializing the DNN, which somehow first learns the speech characteristics and only at a later stage learns how to counteract the adverse acoustic conditions.

  
\section{Conclusions and Future Work} \label{sec:con}
In this paper, some novel methods for taking advantage of contaminated speech training in modern DNN-HMM systems are proposed.
In particular, we explored approaches based on an asymmetric context window, a close-talk supervision and a supervised close-talk pre-training. Experimental results are very promising with a relative performance improvement of more than 15\%  over the baseline system on both real and simulated data. 

Along this line, our next activities will concern the exploration of different architectures, such as convolutional and recurrent NNs.
We also plan to consider different acoustic environments, languages and tasks and, at a later stage, to address a multi-microphone case.

\newpage
\eightpt
\bibliographystyle{IEEEtran}
\bibliography{mybib}

\end{document}